# SAR image matching algorithm based on multi-class features

Zhiqiang Ma, Fengming Zhou

**Abstract:** Synthetic aperture radar has the ability to work 24/7 and 24/7, and has high application value. Propose a new SAR image matching algorithm based on multi class features, mainly using two different types of features: straight lines and regions to enhance the robustness of the matching algorithm; On the basis of using prior knowledge of images, combined with LSD (Line Segment Detector) line detection and template matching algorithm, by analyzing the attribute correlation between line and surface features in SAR images, selecting line and region features in SAR images to match the images, the matching accuracy between SAR images and visible light images is improved, and the probability of matching errors is reduced. The experimental results have verified that this algorithm can obtain high-precision matching results, achieve precise target positioning, and has good robustness to changes in perspective and lighting. The results are accurate and false positives are controllable.

**Keywords:** SAR image match; multi-class features; LSD line detection; template match

## Introduction

Due to the influence of speckle noise and severe local distortion, automated, high-precision, and strong robustness image matching has always been a challenge and bottleneck in the efficient processing and application of SAR images.

Overall, image matching methods are generally divided into grayscale based methods and feature-based methods. The grayscale based method uses the grayscale information of an image or image block, and measures two corresponding images or blocks through similarity measurement algorithms The similarity between image blocks, but due to the high computational complexity of their global optimization, is not suitable for matching with strict time requirements. The feature based approach replaces the analysis of the entire image with the analysis of stable and repetitive image features, reducing computational complexity and sensitivity of matching algorithms to sensor changes; In terms of feature-based matching, an important branch is SAR image matching based on the SIFT framework and its improved models with locally invariant features. Due to the inability of Gaussian filtering to effectively suppress speckle noise and preserve edge or boundary information, the SIFT algorithm and its improved algorithms cannot obtain sufficiently accurate and high-precision feature points in the feature point detection stage, resulting in low matching performance. Based on this, most methods use different Gaussian kernel weighted filters

The method or box filtering method convolves with the original image to construct a scale space to preserve image features at multiple scales, but increases the complexity of the algorithm. Reference [1] points out that during the construction of the scale space stage, Gaussian kernels and their different weighted derivatives are the optimal kernel functions for selection. Reference [2] proposes introducing bilateral filtering into the construction of scale space to preserve more edge information and detect more true feature points. Reference [3] introduces the exponential weighted mean ratio operator into nonlinear diffusion filtering to calculate the gradient ratio and obtain the weight coefficients of the conduction function for SAR images

Speckle noise suppression preserves more edge information but increases time consumption. Reference [4] proposes a fast explicit diffusion algorithm to further reduce the time consumption of solving diffusion equations, but the complexity remains high. [5] Propose using nonlinear diffusion filtering to construct a scale space, and using additive operator splitting algorithm to solve diffusion equations to reduce the time consumption of the scale space. Reference [6] proposes to introduce Gaussian guided filtering into the kernel preprocessing stage of multi-scale space construction at the level of constructing scale space. In the descriptor construction stage, it proposes to use local differential binary algorithm to describe features, reduce the vector dimension of features, and shorten the time for constructing descriptors. Although it is better than the original algorithm in terms of time complexity and matching probability evaluation, it still has a relatively high algorithm complexity overall. Overall, the above bases the matching algorithm of the SIFT framework has two shortcomings: 1) the performance of linear filtering algorithm in preserving edges is low; 2) Nonlinear filtering algorithms are time-consuming and difficult to meet real-time requirements.

Compared to the above matching methods, this article proposes another approach based on feature matching algorithms. For the first time, it combines SAR image line features and regional features to match SAR images, mainly using LSD (Line Segment Detector) line detection algorithm and normalized template matching algorithm. Firstly, fast preprocessing is performed on SAR images, and a custom speckle suppression filter kernel is used to suppress speckle noise in SAR images using their range and azimuth resolutions. This achieves low algorithm complexity while ensuring effective noise suppression; Secondly, combining SAR image prior knowledge (ground distance resolution, azimuth pixel size, and distance pixel size) with the pixel width of visible light images, calculate the scale parameter s between the two; Then, based on line detection, the SAR image is roughly matched to obtain the angle parameters between the SAR image and the visible light image. The first transformation matrix is constructed, and the SAR image is affine transformed using this matrix; Then, the contour moment method is used to filter out small areas, laying the foundation for template matching. The longest straight line feature is then used to locate the template in SAR images, and the normalized template matching algorithm is used to search for the best matching position in visible light images. Finally, a second transformation matrix is constructed by using the coordinates of the midpoint and endpoint of the best matching position and the longest straight line feature, and affine transformation is performed on the SAR image to complete the matching work of the SAR image.

## 1 SAR Image Matching Framework

SAR image matching mainly includes the following parts: speckle noise suppression, matching feature selection, normalized template matching, and matching transformation matrix.

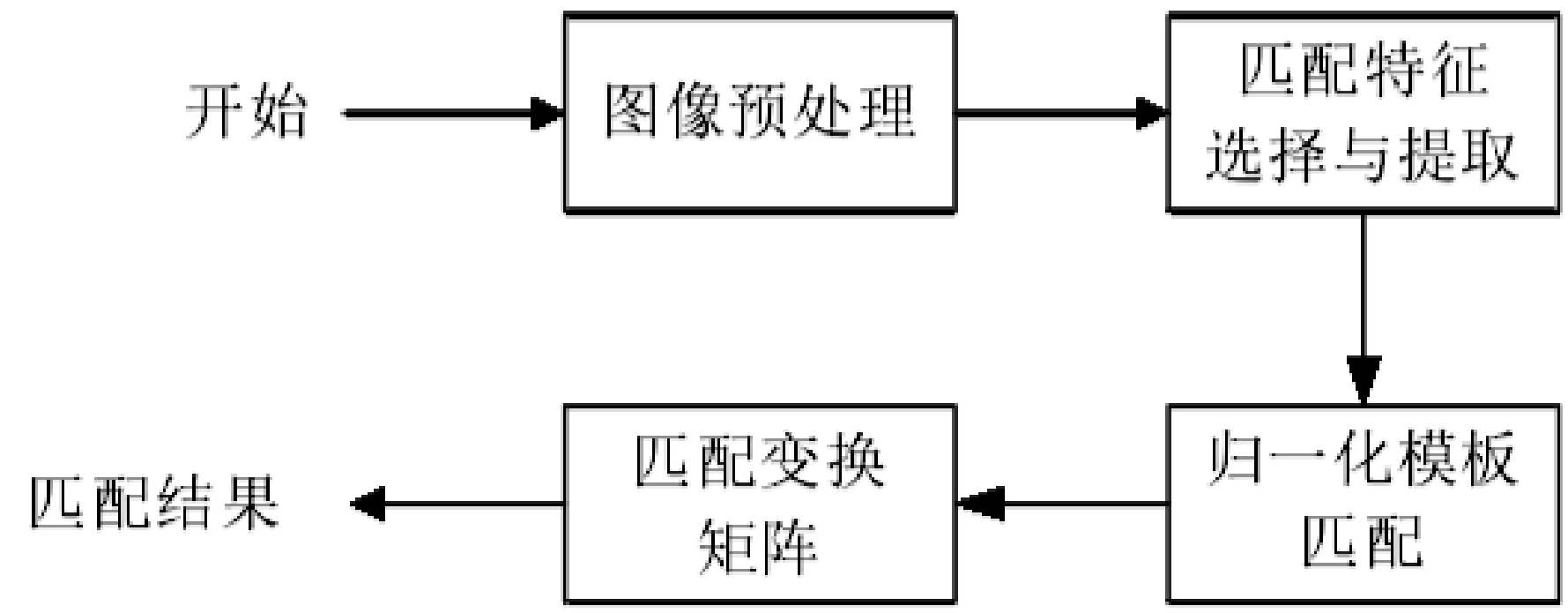

Fig1. SAR Image Matching Framework

SAR image preprocessing mainly includes adaptive binarization and speckle noise suppression. In terms of adaptive binarization threshold design, it can make the features in the binarization image more meet the matching requirements and improve the matching accuracy. Speckle noise suppression is aimed at the multiplicative noise in SAR image. It mainly defines the filter size according to the range resolution and azimuth resolution of SAR image, and then uses the filter for convolution processing according to the distribution attribute of image pixel gray value as the constraint to achieve the effect of noise suppression. In the feature selection module, the line feature and region feature in the image pair are mainly extracted, and the LSD method is used to detect the most representative image

According to the linear features of the image pair, the angle parameters between them can be obtained. In this module, the ground range resolution, azimuth pixel size and range pixel size of SAR image and the pixel width of visible image are used to calculate the scale parameters between them; At the same time, the best matching template in SAR image can be extracted by using line feature. In the normalized template matching module, the matching template of SAR image is used to perform matching operation in the visible image, search for the best matching point pair, and finally calculate the matching change model through the best matching point pair to realize the matching of SAR image.

## 2 SAR image preprocessing

### 2.1 adaptive binarization

Image binarization is an essential operation in the process of image matching. In this paper, a new adaptive threshold binarization method is proposed, which dynamically adjusts the threshold according to the proportion of non-zero pixels in the image, and achieves the binarization result that meets the matching requirements. The main steps are as follows: (1) first, the threshold is given according to the gray attribute of the filtered image, and the preliminary coarse binarization is carried out; (2) The proportion of non-zero pixels of the image in the calculation result; (3) Adjust the binarization threshold according to the proportion; (4) Repeat the above steps until the proportion of non-zero pixels in the image reaches a certain threshold. The effect is shown in the following figure.

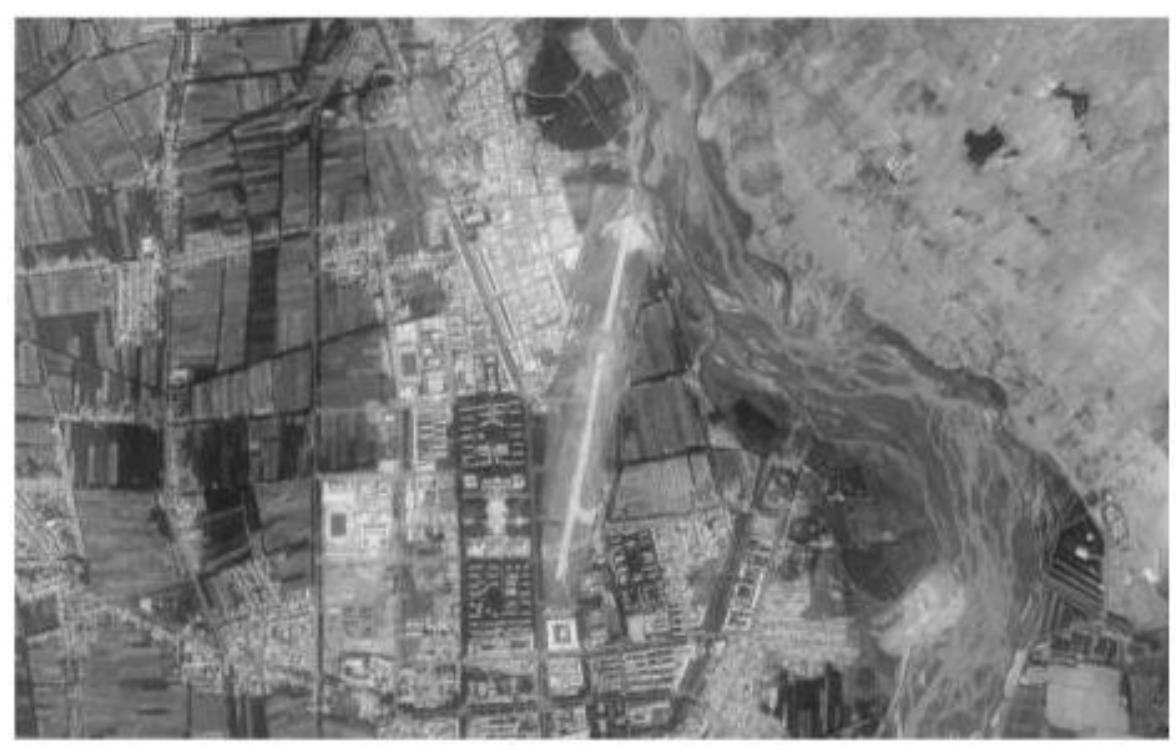

Fig2. Original Image

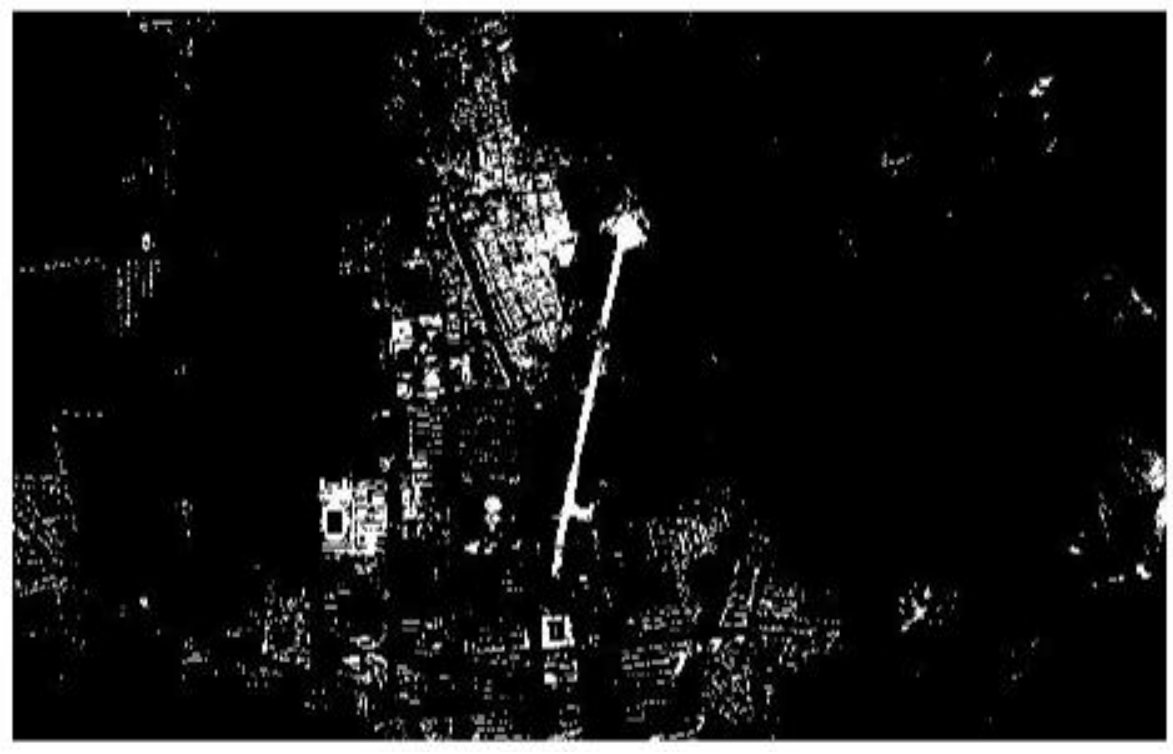

Fig3. Binarize Image

**2.2 speckle suppression**

Due to the inherent speckle noise of SAR image, a large number of matching algorithms can not obtain enough accurate features in the process of feature extraction, resulting in low matching performance and even matching errors. Based on this, this paper constructs a speckle noise suppression filter by using the range resolution DIS and azimuth resolution dre of SAR image. The filter size is defined as (9/DIS) x (9/DRE), and the filter is as follows.

$$kernel(i, j) = \begin{cases} 255, & if[(-9/2Dis)/2 < i < (9/2Dis)/2 \ \&\& \ (9/2Dre)/3 < j < 2*(9/2Dre)/3] \ || \\ & [5*(9/2Dis)/12 < i < 7*(9/2Dis)/12 \ \&\& \ (-9/2Dre)/2 < j < (9/2Dre)/2] \\ 0, & others \end{cases} \tag{1}$$

The specific steps are as follows: (1) the filter is convoluted with the SAR image from left to right and from top to bottom; (2) Judge whether the gray value of the elements in the filter and the corresponding SAR image pixels are equal and equal to 255, and record the number of qualified pixels as 255 num pixel; (3) Judge that the value of the element in the filter is 255, and the gray value of the corresponding binary SAR image pixel is 0, record the number of qualified pixels, and record it as 0/255 num pixel; (4) When the values of 255 num pixel and 0/255 num pixel are greater than 1/3 of the number of filter elements, judge whether the number of pixels with a gray value of 255 in the left and right neighborhoods of each row or the upper and lower neighborhoods of each column in the image exceeds 5, and if so, set the gray value of the target pixel to 0; (5) Repeat the above operation until the filter center position coincides with the pixel in the lower right corner of the image. By using the above noise suppression method, the speckle noise in SAR image can be well suppressed, and the effect is shown in the following figure.

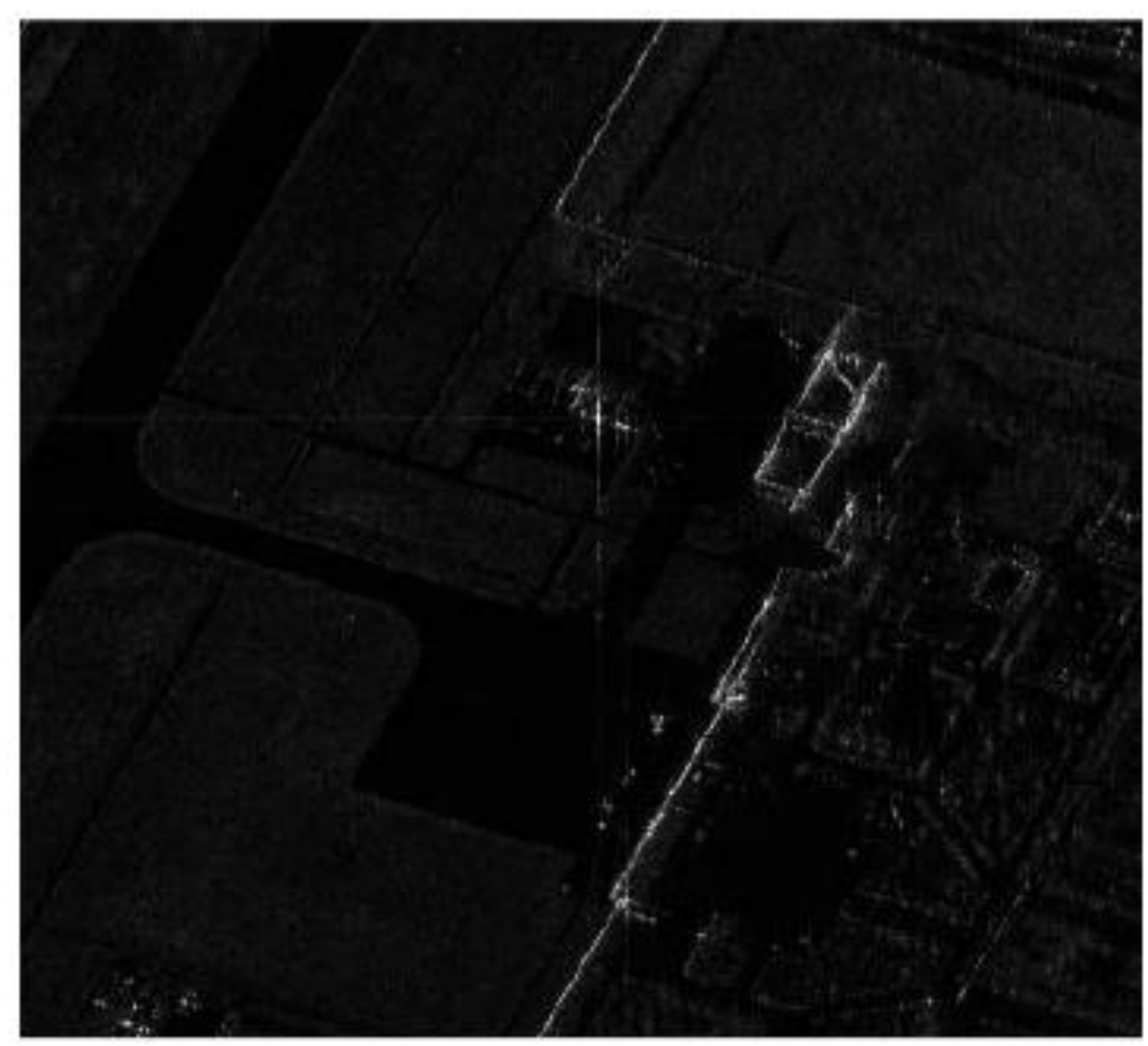

Fig4. Original image to the processed

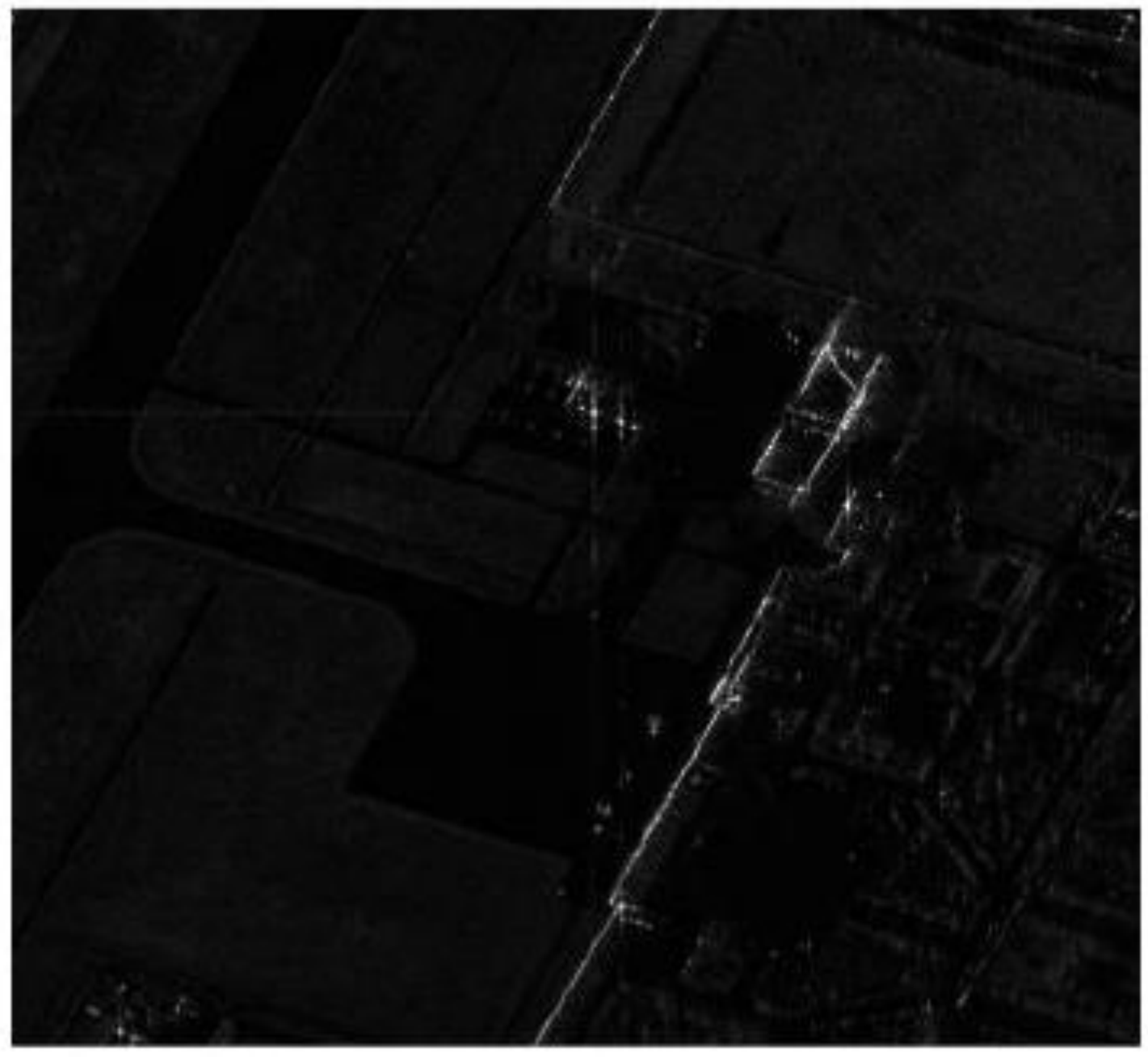

Fig5. Kalman filter renderings

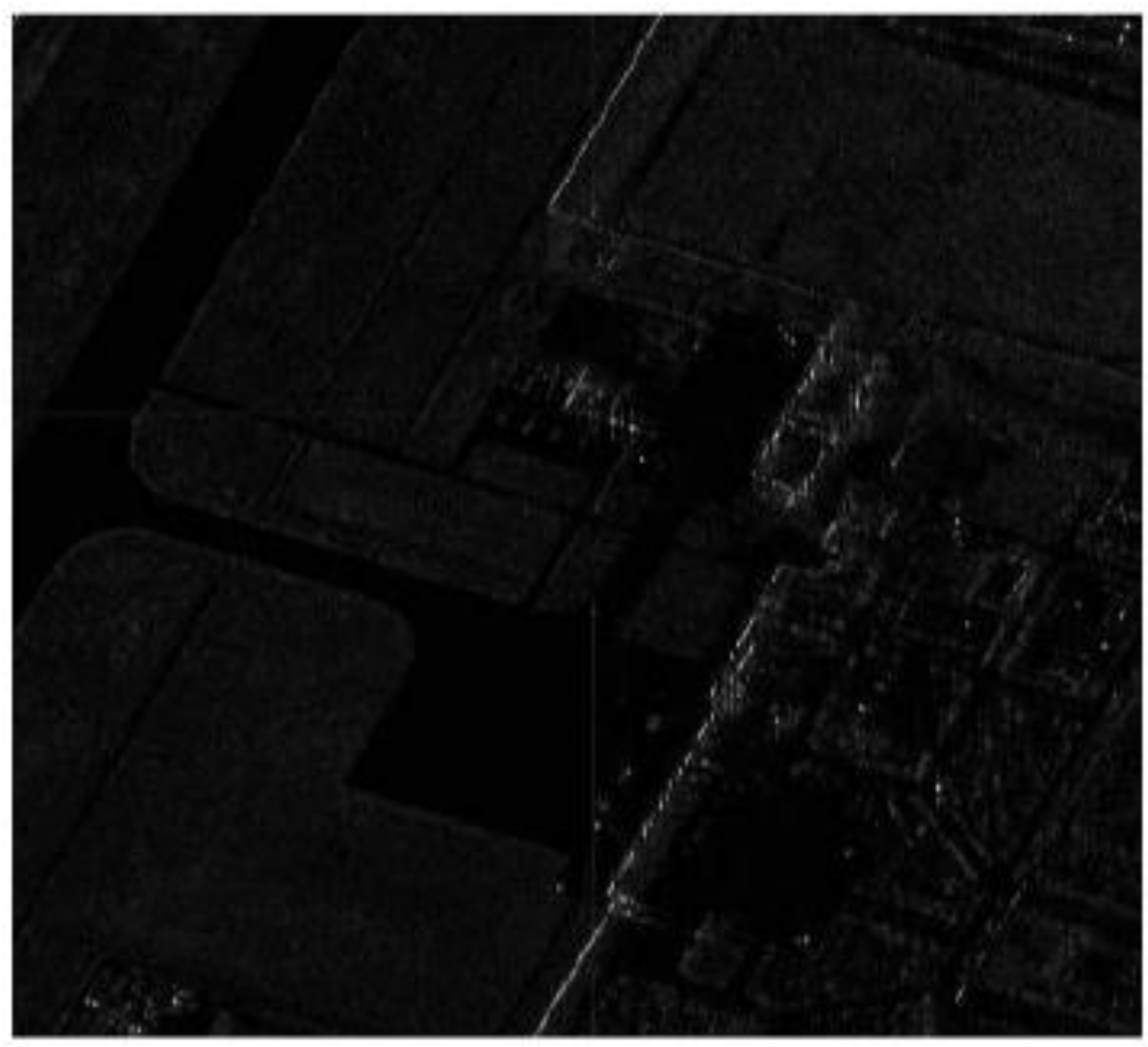

Fig6. The renderings of this method

## 3 image feature extraction and matching

### 3.1 line feature extraction

LSD (line segment detector) is a line detection and segmentation algorithm, which can obtain sub-pixel accuracy detection results in linear time without adjusting any parameters. First, calculate the gradient size and horizontal angle of all pixels in the image, which are recorded as GRD and AGL respectively. According to the gradient size, calculate AGL and gradient amplitude | GRD | as shown in the following formula:

$$\begin{cases} grd_i = \begin{bmatrix} I(i+1, j) + I(i+i, j+1) \\ -I(i, j) - I(i, j+1) \end{bmatrix} / 2 \\ grd_j = \begin{bmatrix} I(i, j+1) + I(i+i, j+1) \\ -I(i, j) - I(i+1, j) \end{bmatrix} / 2 \end{cases} \quad (2)$$

$$agl = \arctan(grd_i / -grd_j) \quad (3)$$

$$| grd(i, j) | = \sqrt{grd_i^2(i, j) + grd_j^2(i, j)} \quad (4)$$

The point with large gradient amplitude is obtained by pseudo sorting as the seed point, and the horizontal line angle AGL of the point is taken as the initial angle of the region. The point whose deviation from the initial angle is less than the preset threshold is found in the eight neighborhoods. The point is added to the region and the initial angle is updated to obtain the line support region. Finally, the line features are obtained by rectangular approximation, and the detection results are arranged in descending order. The first 50 lines are selected as the final detection results to complete the detection of line features.

### 3.1.1 calculation of angle parameters

Calculate the angle parameter theta between the matched images according to the detected line features. The main steps are as follows: first, convert the endpoint coordinates of the line features detected in the SAR image and the visible image into polar coordinates. (1) Calculate the slope k and intercept B of each straight line segment; (2) Calculate the coordinates of the center point of the straight line segment(x_mid,y_mid)、k_mid and b_mid.

$$\begin{cases} x_{mid} = (x_1 + x_2)/2 \\ y_{mid} = (y_1 + y_2)/2 \end{cases} \quad (5)$$

Where $(x_1, y_1)$、$(x_2, y_2)$ is the endpoint coordinate of the straight line segment:

$$k_{mid} = -1/k \quad (6)$$

$$b_{mid} = y_{mid} - k_{mid} \times x_{mid} \quad (7)$$

(3) Calculate the angle of polar coordinates:

When k_mid>0, the following three points can be obtained:

$$O(x_3, y_3) = O(x_{mid}, y_{mid}) \quad (8)$$

$$A(x_4, y_4) = A[x_{mid} - 1, k_{mid} \times (x_{mid} - 1) + b_{mid}] \quad (9)$$

$$B(x_5, y_5) = B(x_{mid} - 1, y_{mid}) \quad (10)$$

When k_mid<0, The following three points can be obtained:

$$O(x_3, y_3) = O(x_{mid}, y_{mid}) \qquad (11)$$

$$A(x_4, y_4) = A[x_{mid} + 1, k_{mid} \times (x_{mid} + 1) + b_{mid}] \qquad (12)$$

$$B(x_5, y_5) = B(x_{mid} - 1, y_{mid}) \qquad (13)$$

You can get the polar angle angle:

$$m_1 = \sqrt{(x_4 - x_3)\times(x_4 - x_3) + (y_4 - y_3)\times(y_4 - y_3)} \qquad (14)$$

$$m_2 = \sqrt{(x_5 - x_3)\times(x_5 - x_3) + (y_5 - y_3)\times(y_5 - y_3)} \qquad (15)$$

$$dot = (x_3 - x_4)\times(x_3 - x_5) + (y_3 - y_4)\times(y_3 - y_5) \qquad (16)$$

$$angle = [(a\cos(dot / (m_1 \times m_2))) \times 180] / 3.1415926 \qquad (17)$$

The second step is to calculate the angle difference between the straight line in the SAR image and the straight line in the visible image based on the visible image.

$$angle_diff = angle_{sar} - angle_{\text{可见光}} \qquad (18)$$

The third step is to count the angle_diff with the same value. The angle parameter theta is the angle_diff with the most occurrences.

**3.1.2 calculation of scale parameters**

In this paper, the scale parameter s between the ground range resolution of SAR image and the pixel width of visible image is calculated. Let the ground range resolution of SAR image be p_sar and the pixel width of visible image be p_w. Calculate the scale parameters from the following formula:

$$s = p_{sar} / p_w \qquad (19)$$

### 3.1.3 SAR image affine transformation

If the width and height of the visible image are w and h respectively, then the coordinate origin can be set as O (w/2, h/2), and the rotation center point of the SAR image can be translated to the origin coordinates. The corresponding transformation matrix is:

$$T = \begin{bmatrix} 1 & 0 & -w/2 \\ 0 & 1 & -h/2 \\ 0 & 0 & 1 \end{bmatrix} \qquad (20)$$

Rotate the SAR image theta around the coordinate origin, and the corresponding transformation matrix is:

$$R = \begin{bmatrix} \cos(theta) & \sin(theta) & 0 \\ -\sin(theta) & \cos(theta) & 0 \\ 0 & 0 & 1 \end{bmatrix} \qquad (21)$$

The image is scaled according to the scale parameter s, and the corresponding transformation matrix is:

$$S = \begin{bmatrix} s & 0 & 0 \\ 0 & s & 0 \\ 0 & 0 & 1 \end{bmatrix} \qquad (22)$$

The affine transformation matrix can be obtained:

$$A = S \times R \times T \qquad (23)$$

According to matrix A, the SAR image can be transformed for the first time, and the orientation and scale of the SAR image can be matched with the visible image to obtain the matching result consistent with the orientation and scale of the visible image, as shown in the following figure.

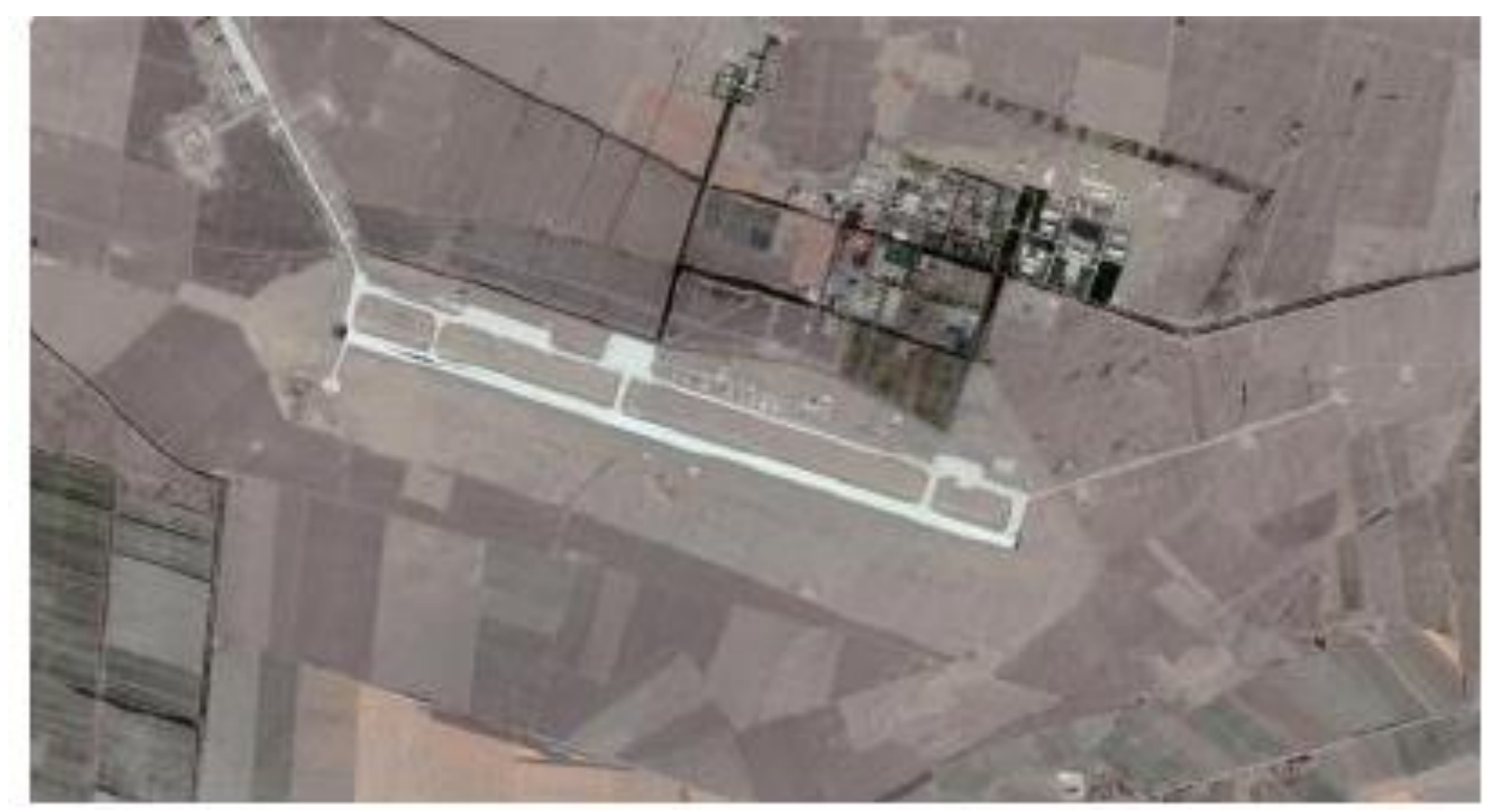

Fig7. Visib le light Image

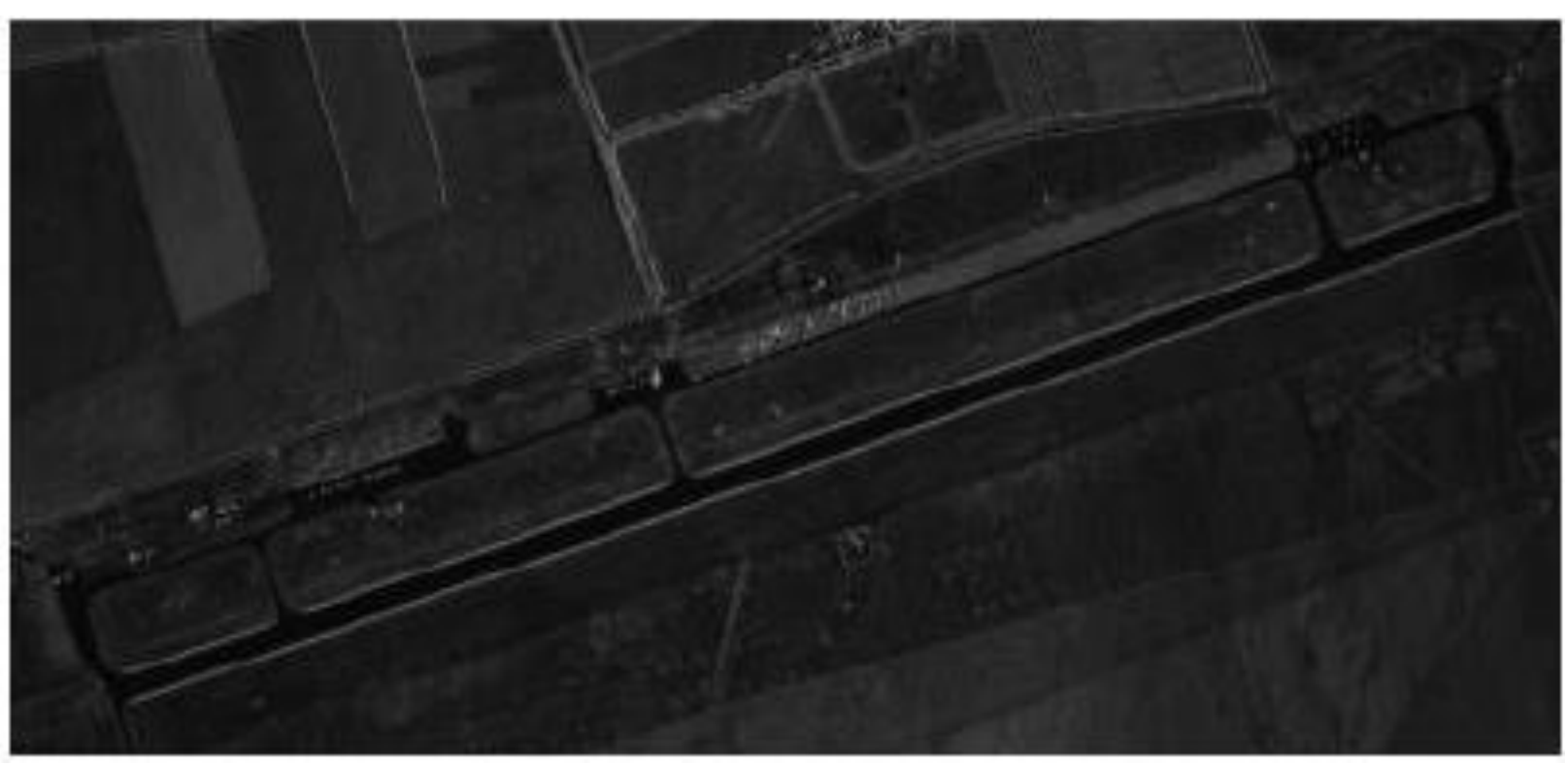

Fig8. SAR Image

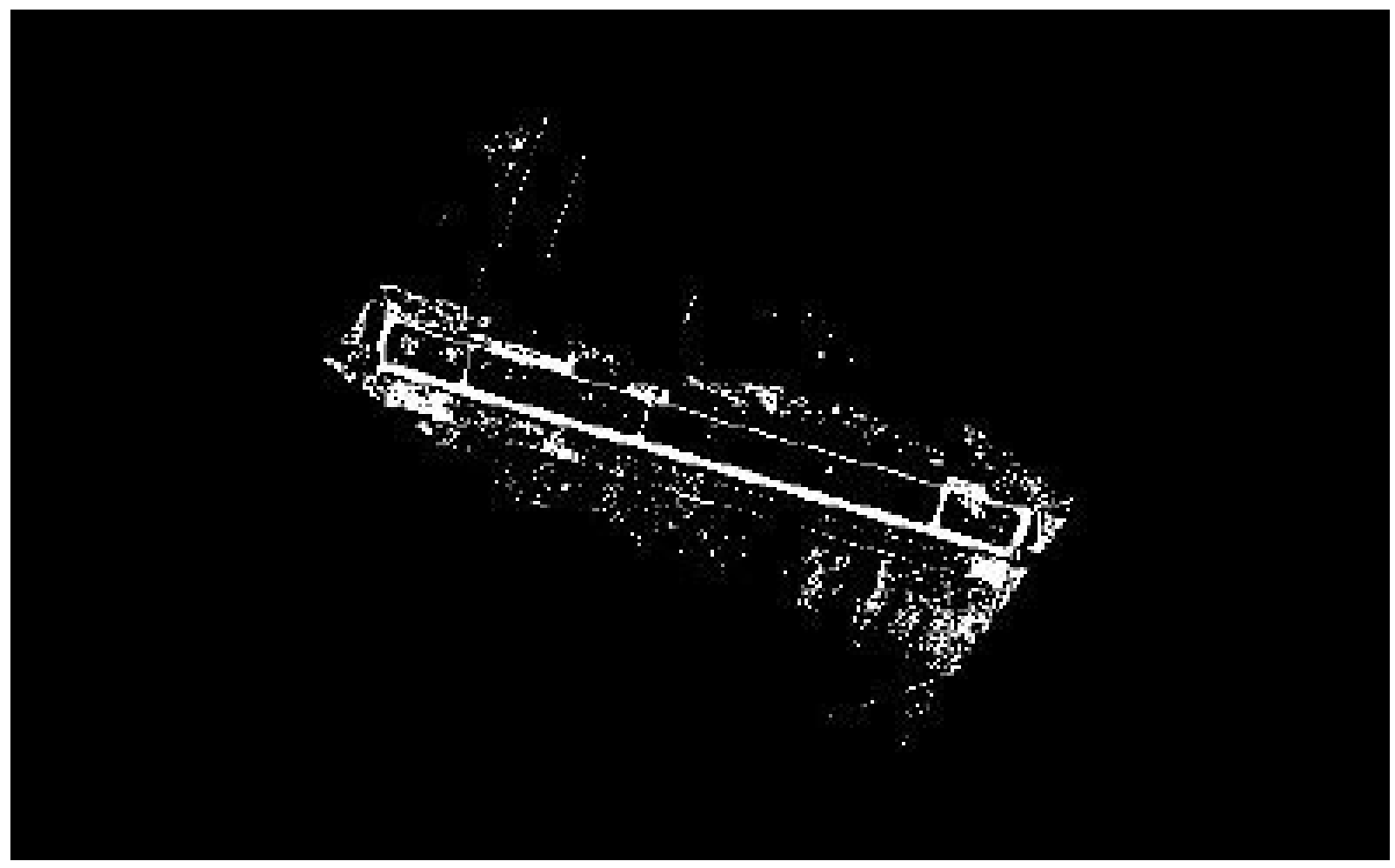

Fig9. Binary image transformation result

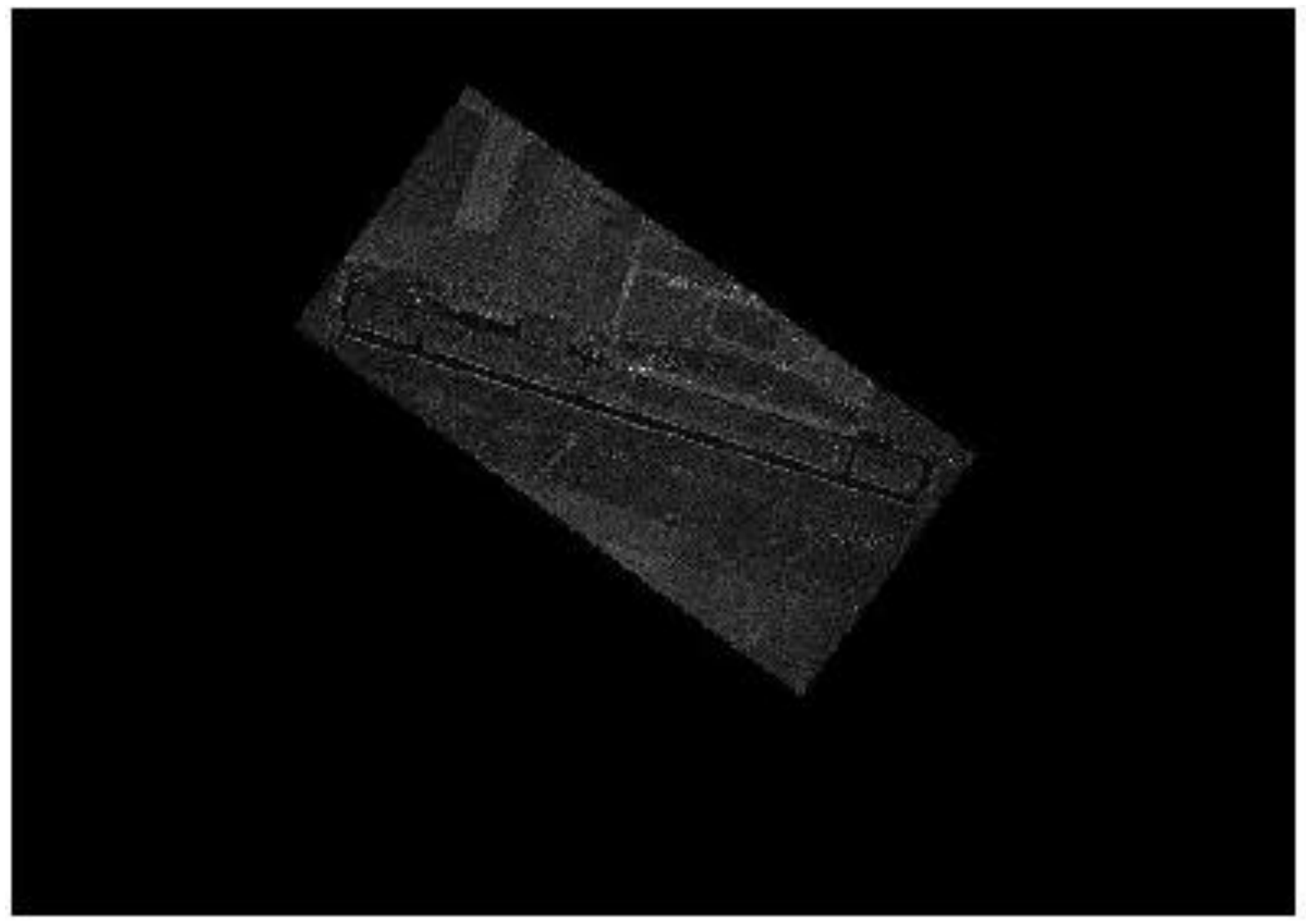

Fig10. Original image transformation result

### 3.2 regional feature extraction

#### 3.2.1 filtering out small areas

In this paper, the contour moment method is used to filter out small and medium-sized areas in the image to improve the subsequent template matching accuracy, as follows:

$$m_{ij} = \sum_{x,y} [C(x, y) \cdot x^i \cdot y^j], i, j = 0,1 \qquad (24)$$

When i, j=0, the area m_00 of the region is obtained. When the value of m_00 is less than a certain threshold, the corresponding region is filtered out, and the effect is shown in the following figure.

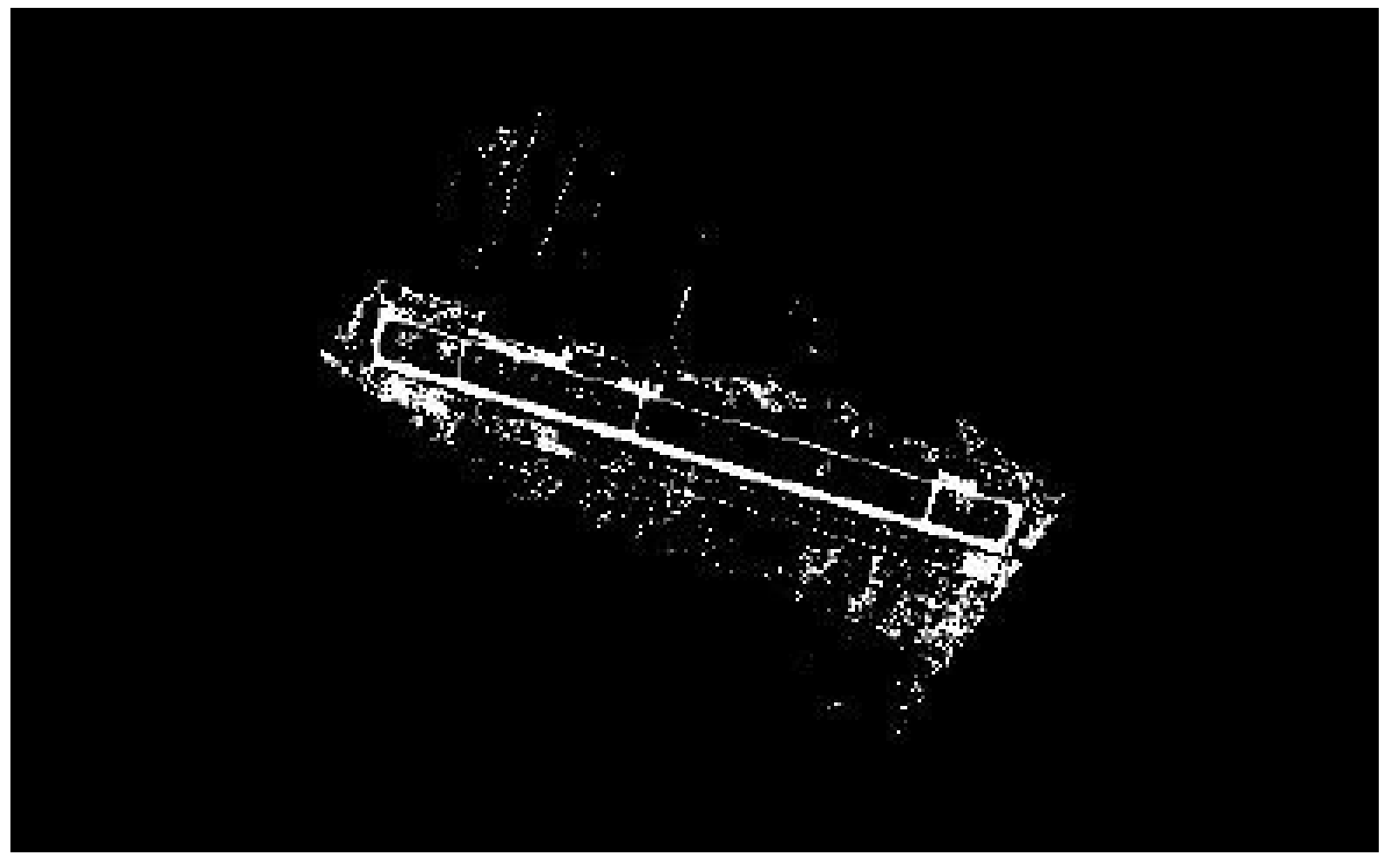

Fig11. Before filtering

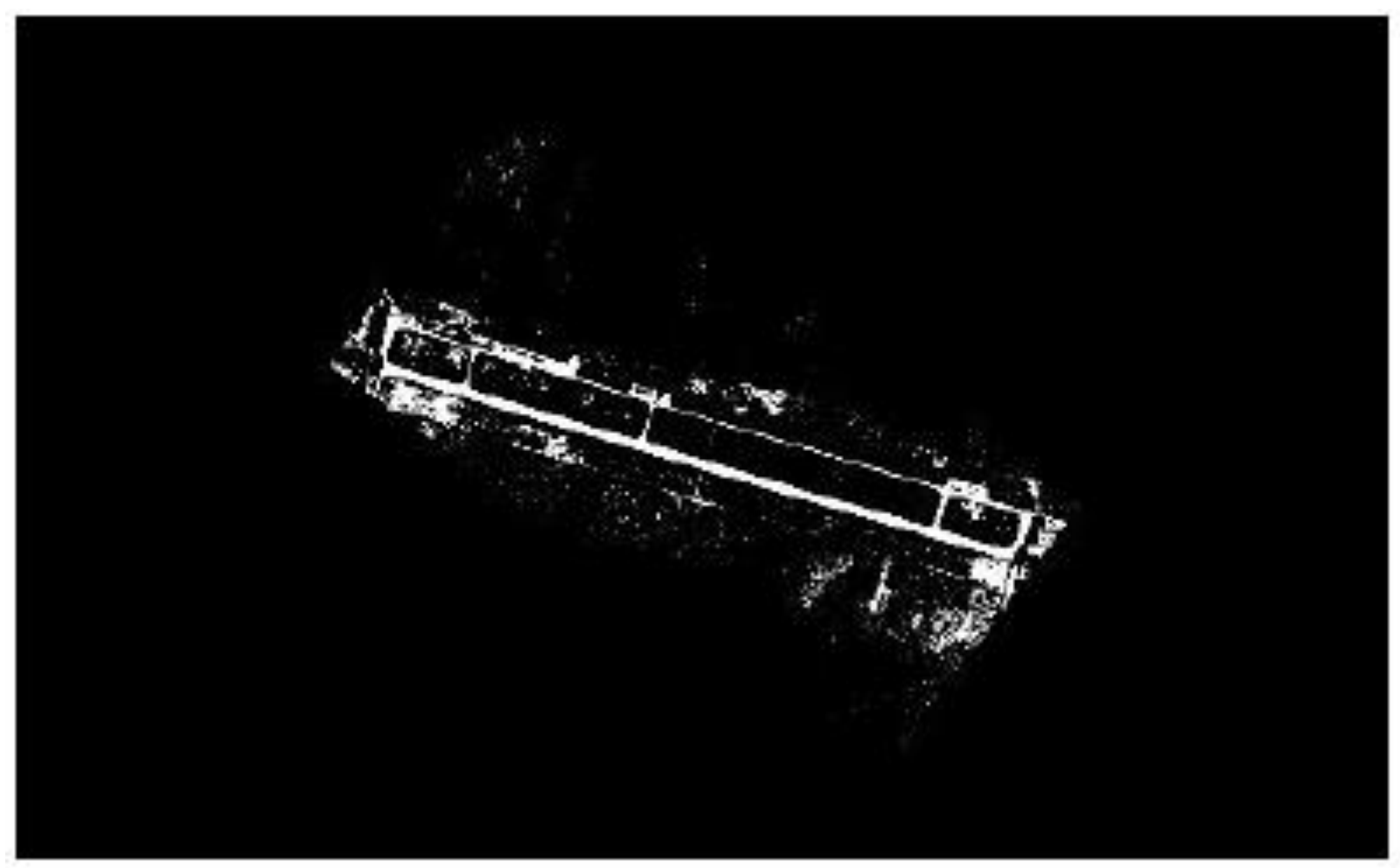

Fig12. After filtering

### 3.2.2 matching template extraction

Using the line feature extraction results, calculate the line length according to the stored line endpoint coordinates, and extract the full the line with sufficient conditions and the longest length, and calculate the midpoint coordinates.

Set the coordinates of the endpoint of the line as $E(x_1, y_1)$ , $F(x_2, y_2)$ Calculate the straight-line length d by using the Euclidean distance formula:

$$d = \sqrt{abs(x_2 - x_1)^2 + abs(y_2 - y_1)^2} \quad (25)$$

Set the midpoint coordinates as $mid(x_{mid}, y_{mid})$ , then:

$$\begin{cases} x_{mid} = (x_1 + x_2)/2 \\ y_{mid} = (y_1 + y_2)/2 \end{cases} \quad (26)$$

Taking the three points $E(x_1, y_1)$ , $F(x_2, y_2)$ , $mid(x_{mid}, y_{mid})$ as the center point, three rectangular regions are extracted from the SAR image, and the rectangular region is the required matching template. The effect is shown in the following figure.

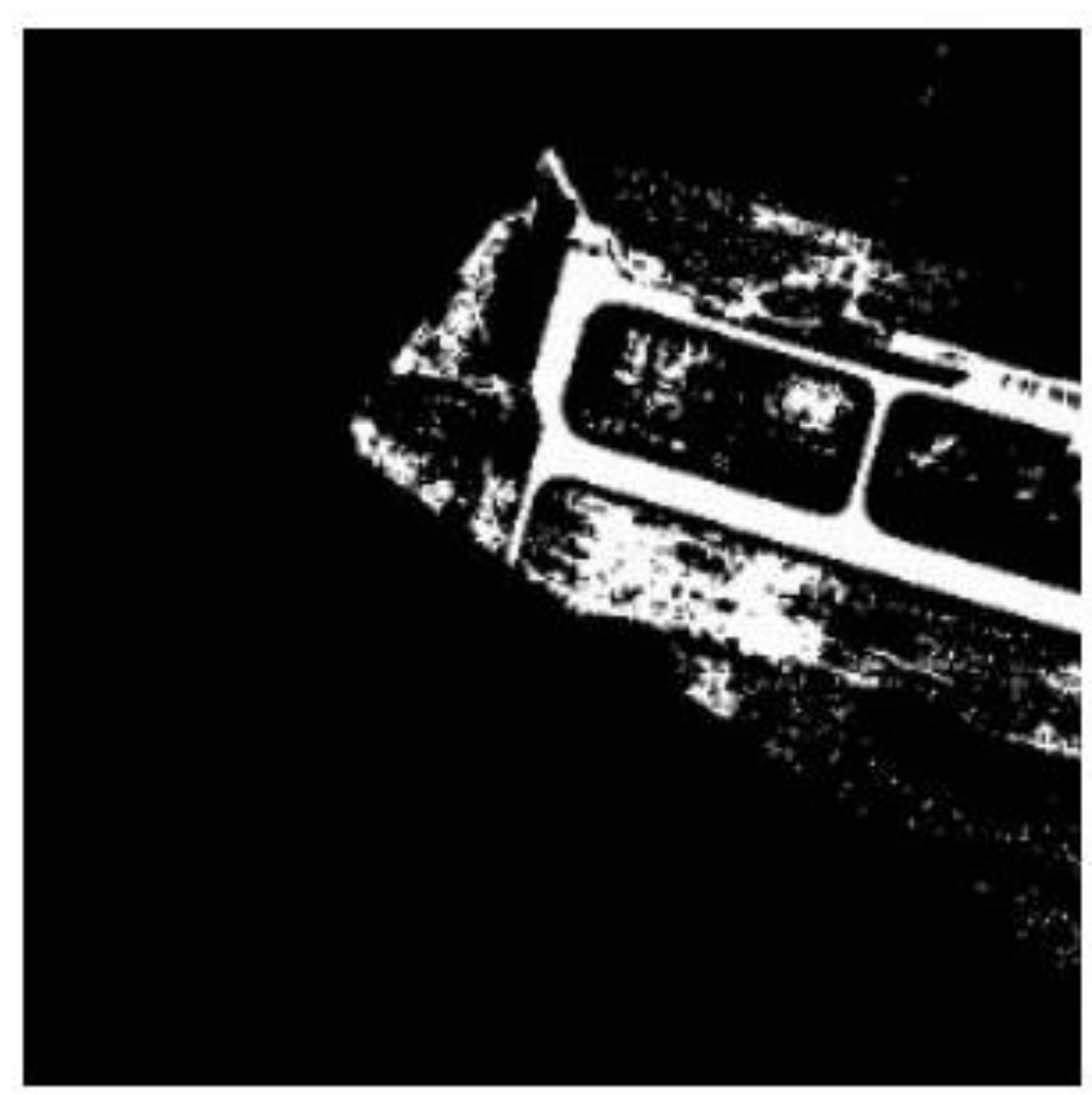

Fig13. Template image 1

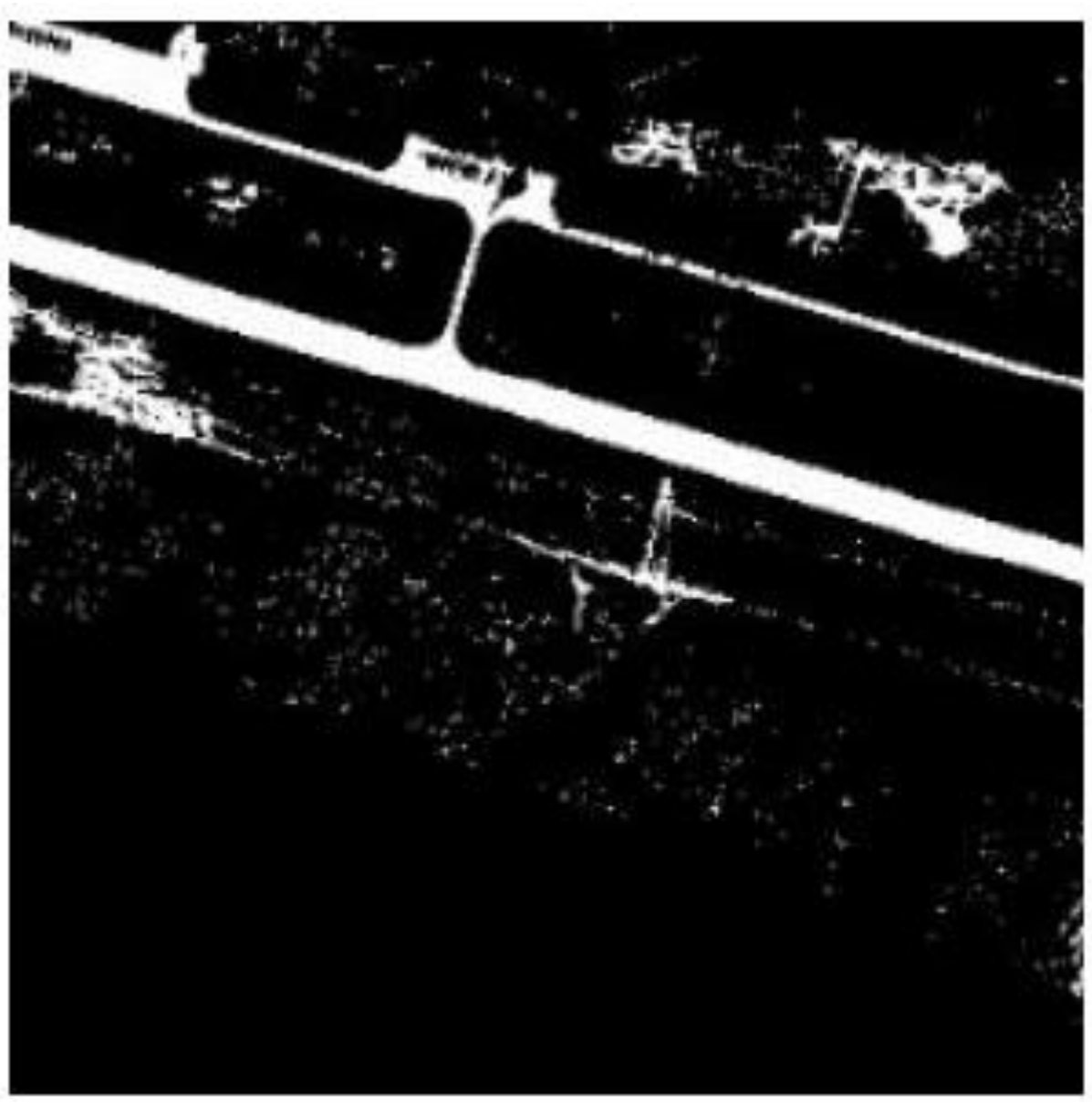

Fig14. Template image 2

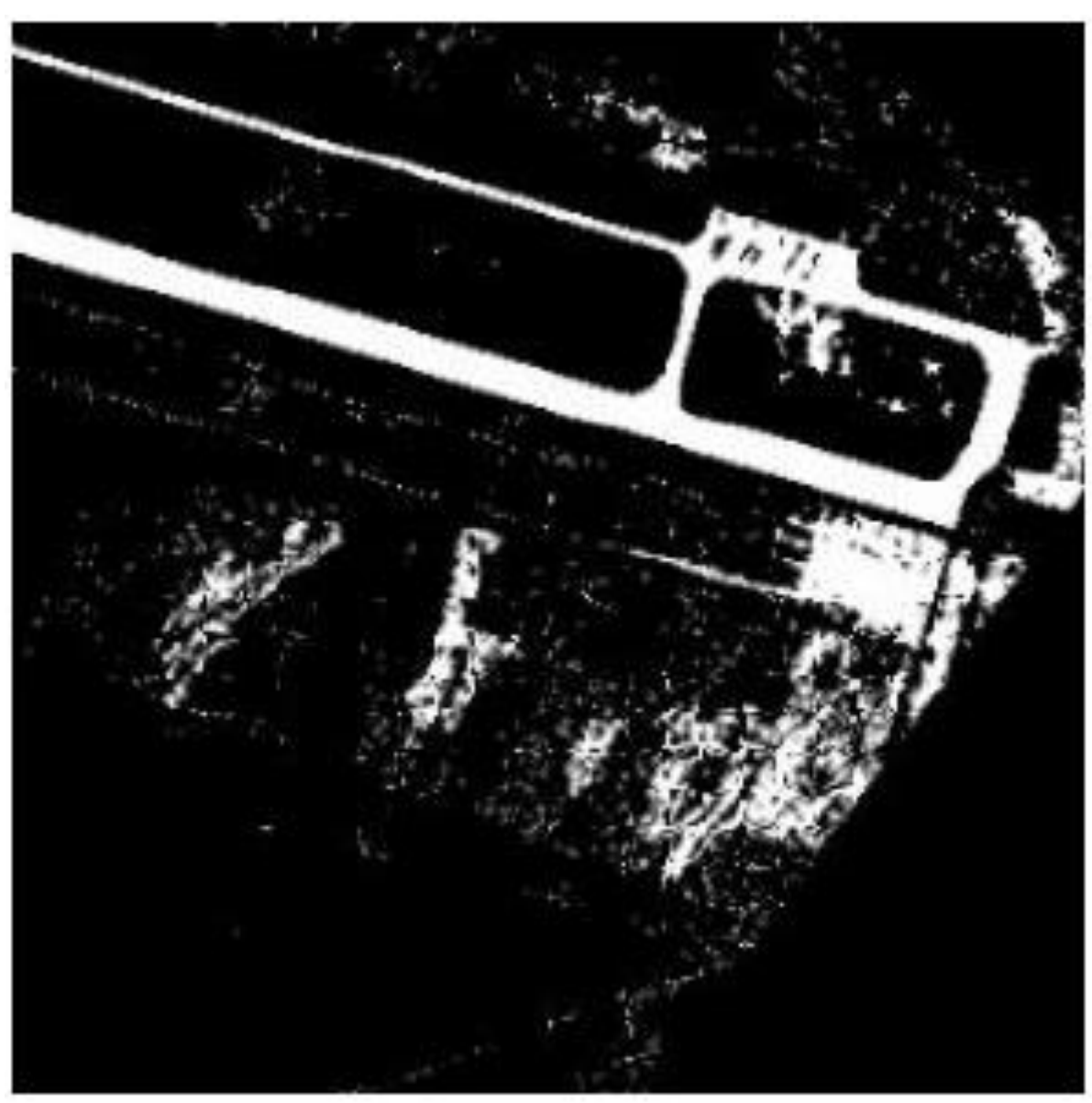

Fig15. Template image 3

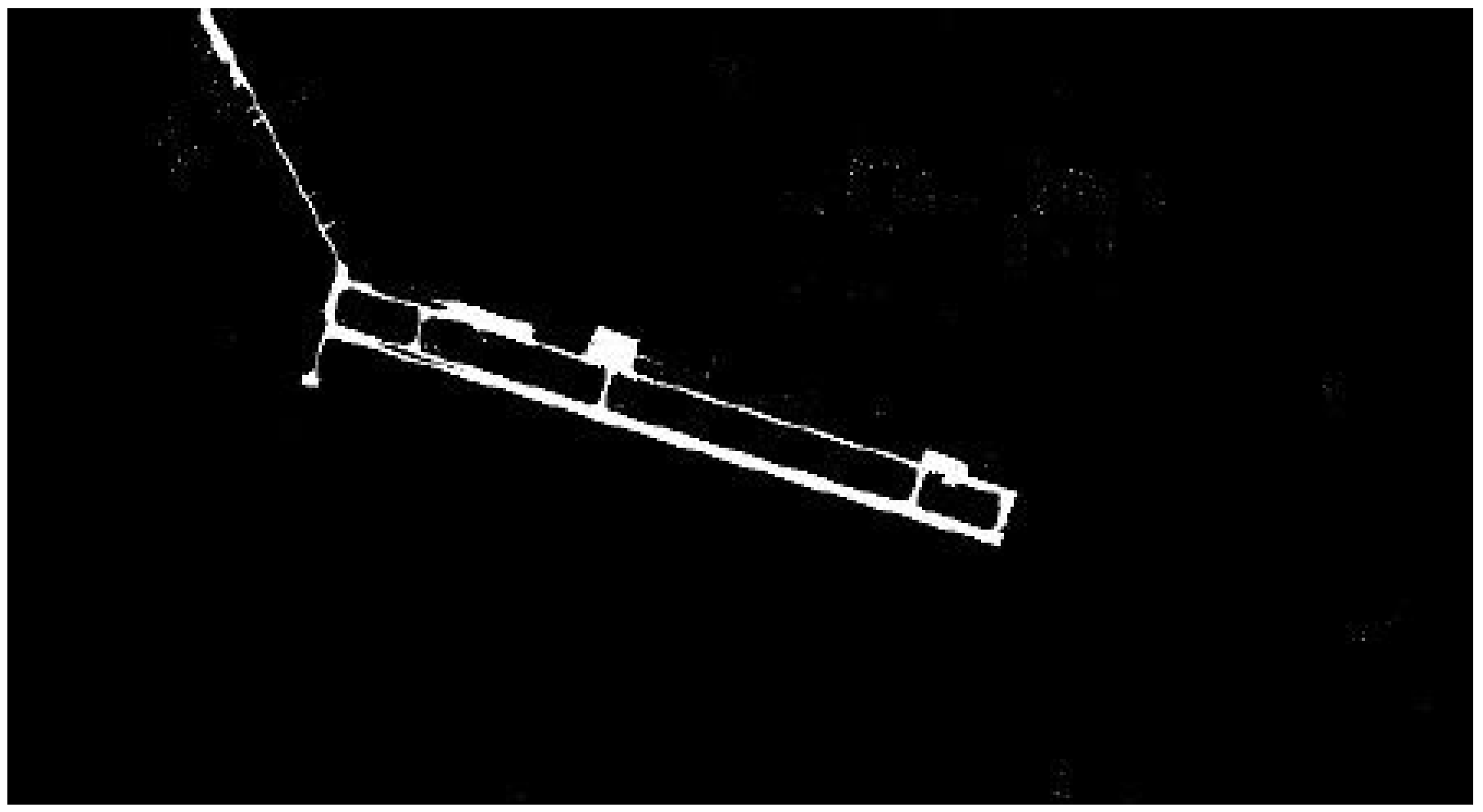

Fig16. The binary result of the visible light image

### 3.3 template matching

#### 3.3.1 calculate the best matching position

Use the above extracted matching template to match the visible image, obtain the coordinate information of the best matching position between the matching template and the visible image, and then use the coordinate pair obtained by matching to calculate the transformation model between the SAR image and the visible image, so as to realize the matching between the SAR image and the visible image. The specific operations are as follows.

In the visible image, slide the matching template from left to right and from top to bottom, and calculate the matching metric R between the template image and the overlapping area. Finally, the result image matrix t_r is obtained. Each position in t_r corresponds to the matching metric of each area in the matching template image and the visible image, where R is calculated by the normalized template matching algorithm, as follows:

$$R(x,y)=\frac{\sum_{x',y'}(T'(x',y')\cdot I'(x+x',y+y'))}{\sqrt{\sum_{x',y'}T'(x',y')^2\cdot\sqrt{\sum_{x',y'}I'(x+x',y+y')^2\cdot}}} \quad (27)$$

In the result image matrix, the position with the largest R value represents the best matching position, and the best matching positions corresponding to the three matching templates are recorded as $(Rx_1,Ry_1)$, $(Rx_2,Ry_2)$, $(Rx_3,Ry_3)$.

**3.3.2 calculation transformation model**

From the above calculated coordinate information, the transformation matrix between SAR image and visible image can be calculated. Let the transformation matrix be:

$$H=\begin{bmatrix} h_{00} & h_{01} & h_{02} \\ h_{10} & h_{11} & h_{12} \end{bmatrix} \quad (28)$$

Let the set of matching points calculated in SAR image be $P_1(x_i,y_i,1)$, where $E(x_1,y_1,1), F(x_2,y_2,1),$ $mid(x_{mid},y_{mid},1)\in P_1$; Let the matching click calculated in the visible light be $P_2(x_i^{'},y_i^{'},1)$, where $(Rx_1,Ry_1,1),(Rx_2,Ry_2,1),(Rx_3,Ry_3,1)\in P_2$ 。. Then there are:

$$P_2^T=HP_1^T \quad (29)$$

The transformation matrix H can be calculated from the above formula. The matching between SAR image and visible image can be realized by using transform matrix H, as shown in the following figure.

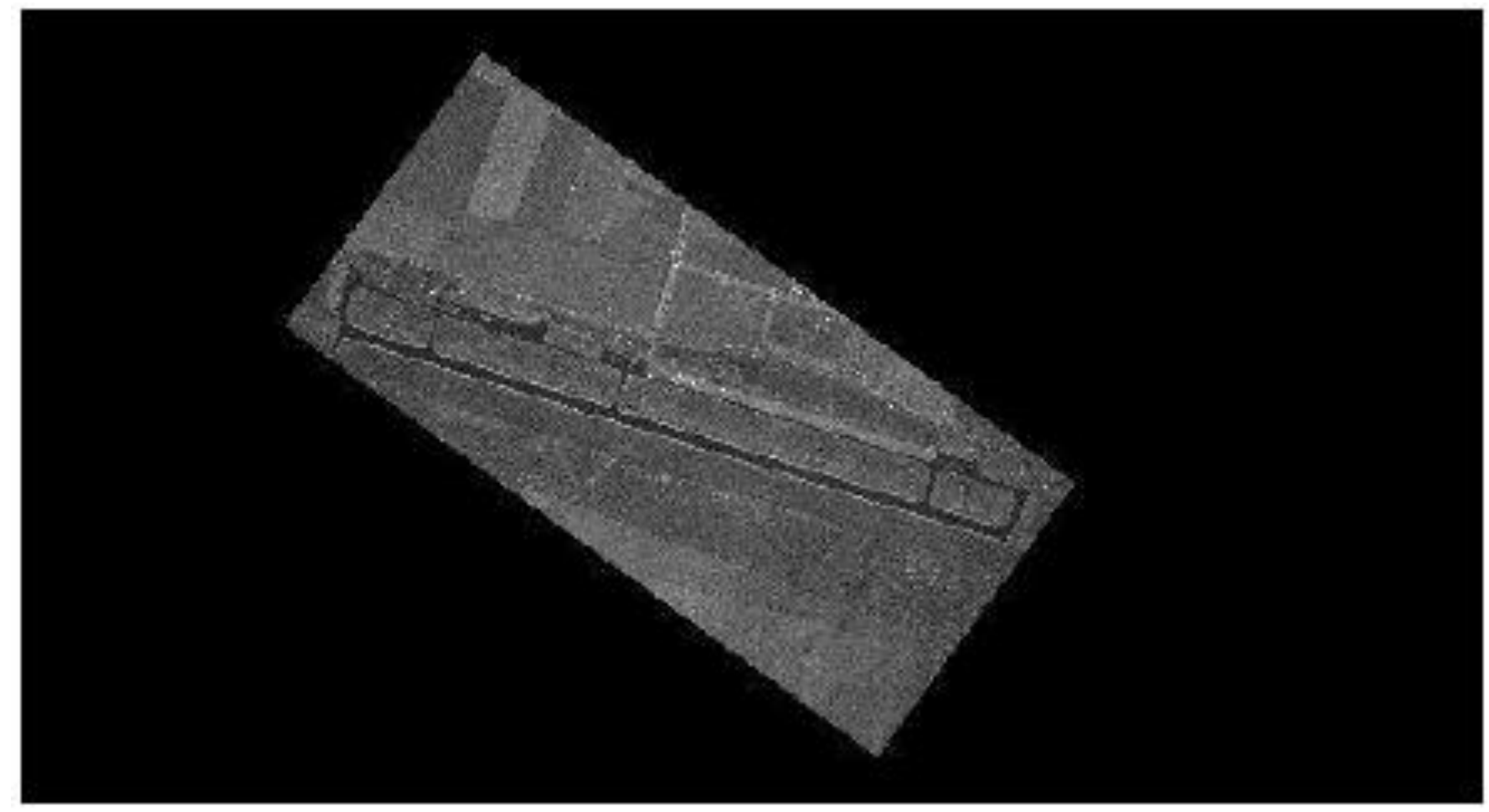

Fig17. The result of matching

## 4 Conclusion

In view of the shortcomings of the matching algorithm based on SIFT framework, this paper proposes a SAR image matching method based on multi class features, which mainly uses two different feature lifting matching algorithms: line and region robustness of; On the basis of image prior knowledge, combined with LSD (line segment detector) line detection and template matching algorithm, through analyzing the attribute association between line and area features of SAR image, the straight line and region feature in SAR image are selected to match the image, which improves the matching accuracy of SAR image and visible image, and reduces the matching error probability. The experimental results verify that the algorithm can obtain high-precision matching results, achieve the accurate positioning of the target, and has good robustness to the changes of viewing angle and illumination, and the results are accurate and the false detection is controllable.